\documentclass[useAMS,usenatbib]{mn2e}
\usepackage{amsmath}
\usepackage{natbib}
\usepackage{graphicx}
\usepackage{color}
\usepackage{rotating}
\usepackage{nicefrac}
\usepackage{txfonts}
\usepackage{subfigure}
\usepackage[percent]{overpic}
\usepackage{xspace}
\usepackage{url}
\usepackage{chemarrow}
\usepackage{amssymb}
\usepackage[percent]{overpic}

\newcommand{\reb}{\texttt{REBOUND}\xspace }
\newcommand{\rev}[1]{#1\xspace }

\title{Large-scale $N$-body simulations of the viscous overstability in Saturn's rings}
\author[Hanno Rein and Henrik N. Latter]{Hanno Rein$^1$ and Henrik N. Latter$^2$\\
\\
$^1$	Institute for Advanced Study, 1 Einstein Drive, Princeton, NJ 08540, e-mail: \url{rein@ias.edu}\\
$^2$    DAMTP, University of Cambridge, CMS, Wilberforce Road, Cambridge CB3 0WA, UK }

\date{Accepted 23 January 2013.  Received 23 January 2013; in original form 28 August 2012.}

\begin{document}
\maketitle
\begin{abstract}
We present results from large-scale particle simulations of the viscous overstability in Saturn's rings.
The overstability generates a variety of structure on scales covering a few hundred metres to several kilometres, including axisymmetric wavetrains and their larger-scale modulations.
Such patterns have been observed in Saturn's rings by the \emph{Cassini} spacecraft.
Our simulations model the collisional evolution of particles in a co-rotating patch of the disk.
These are the largest $N$-body simulations of the viscous overstability yet performed.
The radial box size is five orders of magnitude larger than a typical particle radius, and so describes a 20-50 km radial portion of the rings. 
Its evolution is tracked for more than $10\,000$~orbits.
In agreement with hydrodynamics, our $N$-body simulations reveal that the viscous overstability exhibits a rich set of dynamics characterised by nonlinear travelling waves with wavelengths of a few hundred meters.
In addition, wave defects, such as sources and shocks, punctuate this bed of waves and break them up into large-scale divisions of radial width $\sim 5$ km.
We find that the wavelength of the travelling waves is positively correlated with the mean optical depth.
In order to assess the role of the numerical boundary conditions and also background ring structure, we include simulations of broad spreading rings and simulations with a gradient in the background surface density.
Overall, our numerical results and approach provide a tool with which to interpret \emph{Cassini} occultation observations of microstructure in Saturn's rings.
We present an example of such a synthetic occultation observation and discuss what features to expect.
We make the entire source code freely available.

\end{abstract}


\section{Introduction}
Saturn's rings exhibit a phenomenal wealth of axisymmetric structure covering a vast range of lengthscales, from 100~m to 100~km \rev{\citep{Colwell09}}.
On the shortest scales, ultraviolet and radio occultations reveal that the variations are quasi-periodic with wavelengths between 150~m and 220~m.
These wavetrains intermittently populate regions of optical depth $\sim$ 1-2, such as in the A-ring and less opaque areas in the B-ring \citep{Colwelletal2007,Thomson2007}.
In addition, the \emph{Cassini} cameras have uncovered disjunct bands of varying brightness with widths of 500~m to 10~km \rev{\citep[see Figures 5A and 5F in][]{Porco2005}. 
These intermediate-scale variations differ from the periodic microstructure, however the two can (though not always) coexist \citep{Thomson2007}}.
On the other hand, broad undulations on 100~km scales pervade the inner B-ring, while the central B-ring splits into alternating zones of low and high optical depth, also on scales of 100~km \citep{Colwell09}. 
\rev{This paper addresses the structure on short and intermediate scales (100~m to 10~km). 
We aim to explain why these different features develop and the nature of any relationship between them.}

Fine-scale structure originates in a linear instability connected to the viscous stress of the ring, termed the `viscous overstability'.
Its fastest growing modes take the form of axisymmetric density waves on wavelengths $\sim 50$ m \citep{SchmitTscharn95}.
Such waves give rise to perturbations in the viscous stress that extract energy from the background orbital shear.
This surplus energy is fed into the wave, where it outcompetes standard viscous damping, and thus leads to runaway growth.
In this paper we focus on the small-scale axisymmetric manifestation of the instability, but note that the same mechanism can lead to growth of a global eccentric mode in both narrow and broad rings \citep{Borderies1985,PapaloizouLin1988,LongRap95,Ogilvie2001}.

The viscous overstability has been studied in the context of continuum models (both hydrodynamical and kinetic) and $N$-body simulations, with the early stages of its evolution receiving the greatest attention \citep{SchmitTscharn95,Salo2001,Schmidtetal2001,SaloSchmidtSpahn2001,LatterOgilvie2006a,LatterOgilvie2008}.
The hydrodynamical studies reveal that instability occurs once the ring viscosity increases sufficiently steeply with surface density; i.e.\ if $\beta \equiv d\ln \nu/d\ln \tau \gtrsim 1$, where $\nu$ is viscosity and $\tau$ is normal optical depth.
In kinetic models and $N$-body simulations this instability condition corresponds to $\tau > \tau_c$, where $\tau_c$ is a critical optical depth.
The nonlinear saturation of the instability, on the other hand, has mainly been explored with hydrodynamical models \citep{SchmitTscharn1999,SchmidtSalo2003,LatterOgilvie2009,LatterOgilvie2010}.
The most recent studies show that the long-term evolution of the ring is dominated by nonlinear travelling wavetrains with wavelengths $\sim 200$ m --- shorter waves are vulnerable to secondary instabilities.
The simulated wavetrains undergo small chaotic fluctuations and can be punctuated by wave defects (sources and shocks) that partition the radial domain into structures on larger scales $\gtrsim 1$ km \citep{LatterOgilvie2010}.
The similarity between the shorter wave features and the larger scale divisions, on the one hand, \rev{and the 100~m to 10~km \emph{Cassini} observations}, on the other, is encouraging, though additional work is necessary to make the correspondence tighter.
In particular, the role of self-gravity has yet to be clarified; \rev{it undoubtedly leads to a more complicated set of dynamics than explored here}.

In this paper we simulate the nonlinear saturation of the viscous overstability with large-scale particle simulations in a local co-rotating patch of disk (the shearing box approximation).
Our simulations generalise the previous hydrodynamical results to the more realistic regime of a dense granular flow, in which the ring's pressure tensor deviates significantly from the Newtonian prescription \citep{LatterOgilvie2006a,LatterOgilvie2008}.
They simultaneously extend previous $N$-body simulations to the greater length and time-scale necessary to capture the nonlinear dynamics properly.
These scales are now computationally feasible thanks to the freely available collisional $N$-body \textbf{code} \reb \citep{ReinLiu2012} which includes a symplectic integrator for the shearing sheet and an efficient collision detection algorithm.
We omit self-gravity; its influence will be tested in future work.
A beneficial consequence of this omission is that we can shorten the azimuthal extent of the box.
This permits us to simulate boxes with radial sizes of 55~km~(55\,000~particle radii) for some 10\,000 orbits, orders of magnitudes more than previous simulations.

Our results are in general qualitative agreement with the hydrodynamical simulations of \cite{LatterOgilvie2010}.
Both approaches yield simulations that are dominated by travelling nonlinear wavetrains of comparable morphology and wavelength ($\sim 100$~m).
The particle simulations, however, exhibit a protracted early stage in the evolution in which the counterpropagating waves interact in a complicated way, giving rise to standing wave figures and beating patterns (`wave turbulence').
In addition, the $N$-body simulations yield sources and shock structures as in hydrodynamics though their sizes and morphology differ.
To investigate the potential role of global ring structure (and to minimise that of the periodic boundary conditions), we also performed simulations with viscously spreading rings (mimicking open boundaries) and rings with a background density gradient.
We stress that the structures that emerge in our simulations never exhibit lengthscales longer than roughly 5-10~km.
We conclude that the observed 100~km features in the B-ring are unlikely to be generated by viscous overstability.

New UVIS and VIMS Cassini observations of star occultations have a resolution that is below a few hundred meters such that individual peaks of the overstability-supported wavetrains can be directly observed \rev{\citep{Colwelletal2007,Hedman2012}}.
A direct comparison to numerical simulations \rev{could give constraints} on particles properties such as size, density and coefficient of restitution.
We present the first step towards this goal and present synthetic occultation observations from our
$N$-body simulations that show the feasibility of this approach.

The paper is organised as follows.
In Section~\ref{sec:model} we describe the physical model and governing equations, and then the numerical method we use to solve them.
Section~\ref{sec:tests} provides a brief comparison between our code and the kinetic equilibria and linear theory of \cite{LatterOgilvie2008}, as well as a number of numerical tests.
In Section~\ref{sec:results} we present our main results, beginning with a detailed analysis of a large-scale fiducial simulation and then moving on to the role of the main parameters and global disk structure.
We summarise our results and discuss the implications in Section~\ref{sec:discussion}.

\section{Physical and numerical model}\label{sec:model}
\subsection{Equations of motion} We solve the equations of motion in the Hill approximation \citep{Hill1878} which is a local coordinate system that is co-rotating with a particle on a circular orbit.
The gravity from the central object is linearised in local co-ordinates and the orbital frequency is a constant.
This allows, but does not restrict, us to shear-periodic boundary conditions.
In that case the Hill approximation is also referred to as the shearing sheet.
The $x$, $y$, and $z$ coordinates point in the radial, azimuthal, and vertical direction respectively.
The equations of motion for a test particle can then be written as
\begin{align}
 \ddot x &%
 = 2 \Omega \dot y +3 \Omega^2 x\nonumber\\
 \ddot y &%
 = -2 \Omega \dot x\label{eq:hill}\\
 \ddot z &%
 = - \Omega_z^2 z,\nonumber
\end{align}
where $\Omega$ and $\Omega_z$ are the angular velocity and vertical epicyclic frequency, respectively.
The solution to these equations can be written as epicycles \citep[e.g.][]{ReinTremaine2011}.
In the case where the central object is a point mass and self-gravity is neglected, we have $\Omega=\Omega_z$.
We denote the radial length of the box by $L_x$ and the azimuthal length by $L_y$.
Typically $L_y \ll L_x$ in the simulations we perform.

The only further ingredients needed besides Eqs.~(\ref{eq:hill}) are the finite particle \rev{radius} $r_p$ and a collision model.
We treat particles as hard spheres (they are not permitted to overlap) and the outcome of a collision is described using a normal coefficient of restitution $\epsilon$, which can be either constant or a function of the impact velocity $v_\text{imp}$.
The particles have no spin.

In this paper we do not treat self-gravity directly.
However, we increase the vertical epicyclic frequency in some simulations $\Omega_z = 3.6\Omega$ in order to concentrate particles in the mid-plane, thus mimicking self-gravity's vertical compression \rev{and consequent enhancement of the collision frequency}.
This convenient method has been used in previous ring studies \citep[e.g.][]{Wisdom1988,SaloSchmidtSpahn2001,SchmidtSalo2003}.
It allows us to remove one scale from the problem (i.e.\ the particle density or Hill radius) and more easily focus on the fundamental details of the collective axisymmetric dynamics.

We plan to study simulations including self-gravity in a future paper.
Self-gravity is important because it may instigate non-axisymmetric wake structures that could compete with the overstability and alter its saturation \citep{SaloSchmidtSpahn2001}.
Self-gravity will furthermore change the properties of the overstability waves directly through their nonlinear dispersion relation.

\subsection{Diagnostics} In order to probe the collective behaviour of the granular flow, we require a number of averaged quantities.
We define the mean normal geometrical optical depth $\bar\tau$ as the total projected area of the particles on the $(x,y)$ plane divided by the total area of the $(x,y)$ plane.
In other words
\begin{equation}
 \bar\tau = N\,\pi\,r_p^2/(L_x L_y),
\end{equation}
where $N$ is the number of particles.
Thus, $\bar\tau$ is stipulated at the beginning of each run and will not change.

We also define the radially and temporally varying optical depth, denoted simply by $\tau$, by subdividing the radial domain into thin strips of radial length $L_S$:
\begin{equation}
 \tau(x_i,\,t) = N_i(t)\,\pi\,r_p^2/(L_S L_y), 
\end{equation}
where $x_i$ is the radial location and $N_i(t)$ is the number of particles in the $i$'th strip \rev{at time $t$.}
As we typically begin a run from a homogeneous state, at $t=0$ we have $\tau(x_i)=\bar\tau$ for all $i$.
The ensuing overstable fluctuations will subsequently be captured by the variations in $\tau(x_i)$.
$L_S$ is chosen to be smaller than any scale we are interested in but not so small as to be influenced by Poisson noise.

We further introduce the photometric optical depth $\hat\tau$ which we define as one minus the transmission.
The transmission is the fraction of light passing through the ring, \rev{and must be} 1 for $\hat\tau = 0$ and 0 for $\hat\tau=1$.
This is calculated with a Monte Carlo Ray tracing algorithm.
The algorithm shoots rays through a random point in the simulation box but with a predefined direction (the viewing angle).
Then, we check for intersections with ring particles along the ray.
This is repeated until a desired accuracy has been reached.
Unless otherwise noted we use a viewing angle normal to the ring plane.

The filling factor is defined as the proportion of volume taken up by the particles.
For spherical particles it can be defined as $FF=(4\pi/3)n r_p^3$, where $n$ is volumetric number density.
Particularly useful is the filling factor at the midplane $FF_0$, which requires the calculation of the number density at $z=0$.

Let us further introduce the mean velocity dispersion tensor which is computed from 
\begin{equation}
 W_{ij} = \langle \dot{x}_i\dot{x}_j \rangle,
\end{equation}
\rev{where $(\dot x_1,\dot x_2,\dot x_3)=(\dot x,\dot y +\frac32 \Omega x ,\dot z)$ is the velocity relative to the shear} and the angle brackets indicate a suitable average over the particles and possibly over time.
The velocity dispersion $c^2$ is then $W_{ii}/3$.  
The translational (local) component of the kinematic viscosity is simply
\begin{equation}
 \nu_\text{trans} = \frac23 W_{xy}/\Omega.
\end{equation}
The collisional (nonlocal) component of the viscosity is 
\begin{equation}
 \nu_\text{coll} = \frac{2}{3\Omega\,M\,T}\sum (x_> - x_<)\Delta p_y,
\end{equation}
where the sum is taken over all binary collisions that occur in a time interval $T$.
Here $M$ is the total mass of all ring particles, $\Delta p_y$ denotes the transfer of $y$ momentum from the inner to the outer particle in such a collision, and $x_>$ and $x_<$ the radial locations of the two impacting particles \citep{Wisdom1988, Daisaka2001}.
As we neglect self-gravity there is no gravitational or wake contribution to the overall momentum transport.

\subsection{N-body code and algorithms used} 
\begin{table*}
 \centering
 \begin{tabular}{l|llllllll }
  \hline Name / Description &
  $r_p$ [m] &
  $L_x$ [m] &
  $L_y$ [m] &
  $\bar\tau$&
  $N$ &
  $\Omega_z
  [\Omega]$ &
  $\epsilon$\\
  \hline \hline equilibrium			& 1 	& 25			& 25		& 0.1 - 3			& 20 - 600	& 1	& 0.5\\
  convergence			& 1 	& 32 - 8192 (1024)	& 2 - 40 (5)	& 1.64				& 32 - 21\,382	& 3.6	& 0.5\\
  fiducial 			& 1	& 18\,641		& 15.53 	& 1.64 			& 151\,179 	& 3.6	& 0.5\\
  varying tau (probing wavelength dependence) 	 			& 1	& 3\,976 		& 15.53 	& 0.1 - 2.7			& 1\,967 - 53\,097	 	& 3.6	& 0.5\\
  very low optical depth 		& 1	& 18\,641		& 15.53 	& 0.50				& 46\,092	& 3.6	& 0.5\\
  low optical depth 		& 1	& 18\,641		& 15.53 	& 1.00 			& 92\,183	& 3.6	& 0.5\\
  high optical depth	 	& 1	& 18\,641		& 15.53 	& 2.00 			& 184\,365	& 3.6	& 0.5\\
  density ramp 			& 1	& 18\,641		& 15.53 	& 0.04 - 3.24 (left - right) 	& 143\,070	& 3.6	& 0.5\\
  open boundaries			& 1	& 16\,641		& 15.53 	& 1.64 			& 134\,960 	& 3.6	& 0.5\\
  fiducial super wide		& 1	& 55\,922		& 5.17	 	& 1.64 			& 151\,162 	& 3.6	& 0.5\\
  size distribution		& 0.5-2	& 18\,641		& 15.53 	& 1.64 			& 204\,178 	& 3.6	& 0.5\\
  inelastic collisions		& 1	& 18\,641		& 15.53 	& 1.64 			& 151\,179 	& 3.6	& 0.0\\
  \cite{Bridges1984} collision law 				& 1	& 18\,641		& 15.53 	& 1.64 			& 151\,179 	& 3.6	& var.\\
  \hline \end{tabular}
 \caption{\label{table:simulationlist} Parameters of simulations presented in this paper.
  See text for details.}
\end{table*}

We use the freely available $N$-body code \reb \citep{ReinLiu2012} to perform all of the simulations presented in the paper.
The code is ideally suited for this kind of study because of both speed and accuracy.
We use the highly efficient and accurate mixed variable symplectic integrator SEI \citep{ReinTremaine2011}.
Collisions are detected using a so called plane-sweep algorithm.
The plane-sweep algorithm scales linearly with the number of particles,
$O(N)$, for highly elongated boxes such as those considered here.
Both algorithms are already implemented in \reb.
These advantages allow us to use boxes which are
$10^5$~times larger than the typical particle radius.
We therefore do not need to rely on periodic boundary conditions and can study effects near ring boundaries and density gradients.

\rev{In early test simulations we noticed that it was crucial not to introduce a preferred direction in the numerical scheme.
The symplectic integrator introduced by \cite{ReinTremaine2011} ensures that important symmetries are not broken, even on the scale where machine precision becomes important.
However, collisions formally break the symplectic nature of our implementation.
We therefore chose to randomise the order in which collisions are resolved after each timestep.
This removes all spurious correlations which might otherwise be introduced when choosing a specific order in which collisions are resolved (i.e. resolving them from left to right, by a numerical particle identifier or by the position in memory).
Unfortunately, this complicates the parallelisation on distributed memory systems.
In the production runs presented in this paper, we were not able to reproduce a dependence on the order in which the collisions are resolved.
However, to be absolutely safe, we decided to keep the randomisation turned on.}

For the reader's convenience, we list all the simulations presented in this paper in Table~\ref{table:simulationlist}.
The first column describes the simulation.
The second column lists the particle radius.
The third and fourth columns list the radial and azimuthal size of the box.
The fifth and sixth columns list the initial mean optical depth
$\bar\tau$ and the number of particles.
The seventh column lists the vertical epicyclic frequency in units of the standard epicyclic frequency.
The eighth column lists the coefficient of restitution.

\section{Comparisons and numerical tests}\label{sec:tests}
\subsection{Background equilibrium state}
\begin{figure*}
 \centering \resizebox{0.99\textwidth}{!}{\includegraphics{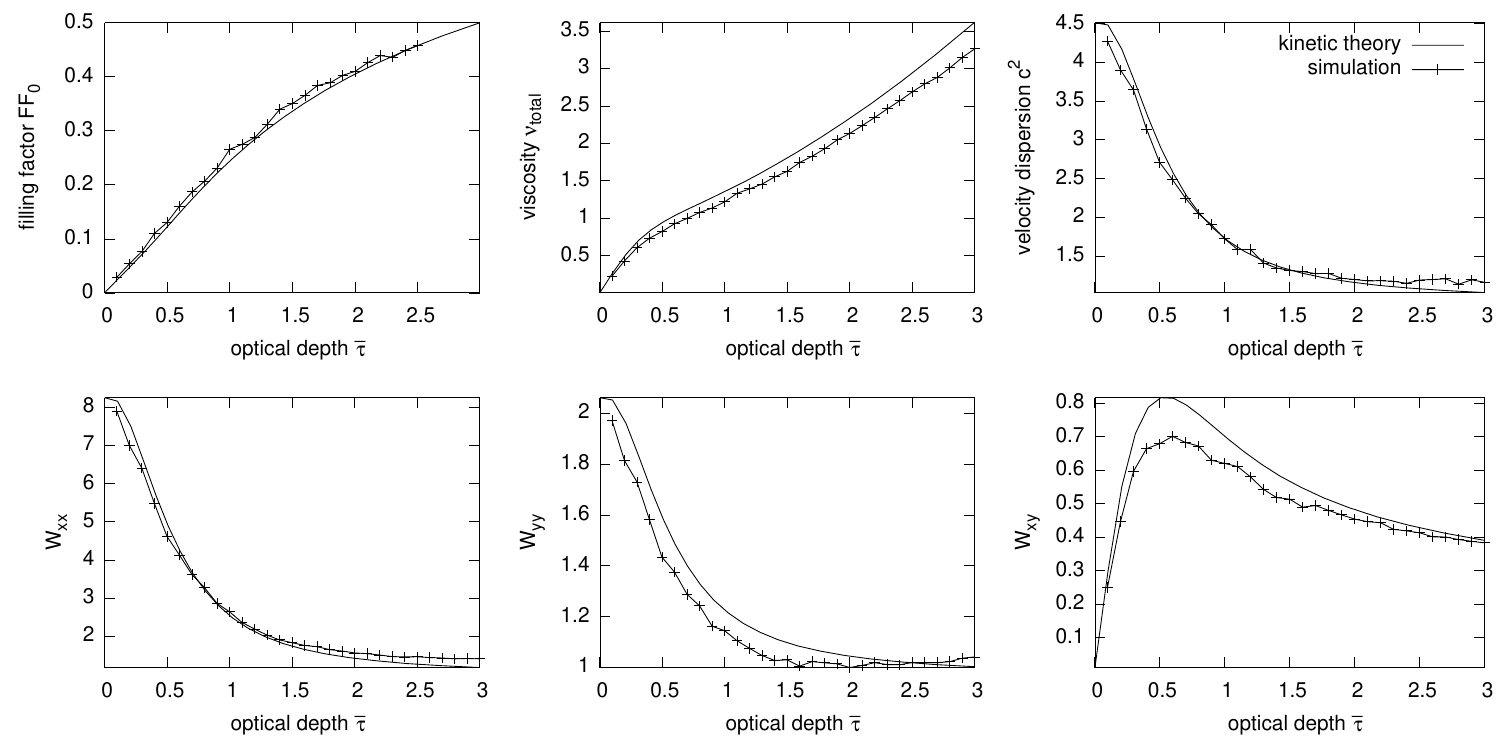}}
 \caption{Comparison of numerical simulations to kinetic theory.
  From top left to bottom right: filling factor, total viscosity, velocity dispersion, $xx$, $yy$ and $xy$ components of the velocity dispersion tensor.
  \rev{We use dimensionless units where time is measured in terms of $\Omega^{-1}$ and distance in terms of the particle radius $r_p$.}
  See text for details.\label{fig:comparison}}
\end{figure*}

We begin by calculating the properties of homogenous statistical equilibria with our numerical code and comparing them to the kinetic theory of \cite{ArakiTremaine1986} as reformulated by \cite{LatterOgilvie2008}.
We use a simulation box with a width of only 25 particle radii.
As a consequence, growing overstability modes cannot fit into the domain (see also Sect.~\ref{sec:convergence}).
Depending on the optical depth, the number of particles ranges from $N=20$~to~$600$.
A constant coefficient of restitution $\epsilon=0.5$ is used.
Other than the particle radius $r_p$ and the frequencies $\Omega$ and $\Omega_z$, which we set equal to 1 in our units, there is no other scale.

In Figure~\ref{fig:comparison} we plot the filling factor in the midplane $FF_0$, the velocity dispersion $c^2$, the $xx$, $yy$ and $xy$ components of the velocity dispersion tensor $W_{ij}$, and the total kinematic viscosity $\nu_\text{total}=\nu_\text{coll}+\nu_\text{trans}$.
These quantities are plotted as functions of the optical depth $\bar\tau$.
Overall, the numerical equilibria are in good agreement with the kinetic theory, with errors of similar order \rev{to} previous comparisons \citep{Wisdom1988,LatterOgilvie2008}.
At larger $\bar\tau$, systematic discrepancies emerge which are the result of the large filling factors $FF_0\approx 0.4$.
In this regime the Enskog approximation begins to break down, and the kinetic results are less reliable.
There is also a 10\% error in the $xy$ component of the viscous stress at lower $\bar\tau$, which has been previously noted \citep{LatterOgilvie2008}. 
These results show that our particle code correctly reproduces established kinetic ring equilibria.

\subsection{Linear stability}\label{sec:array}
\begin{figure}
 \centering \resizebox{0.99\columnwidth}{!}{\includegraphics{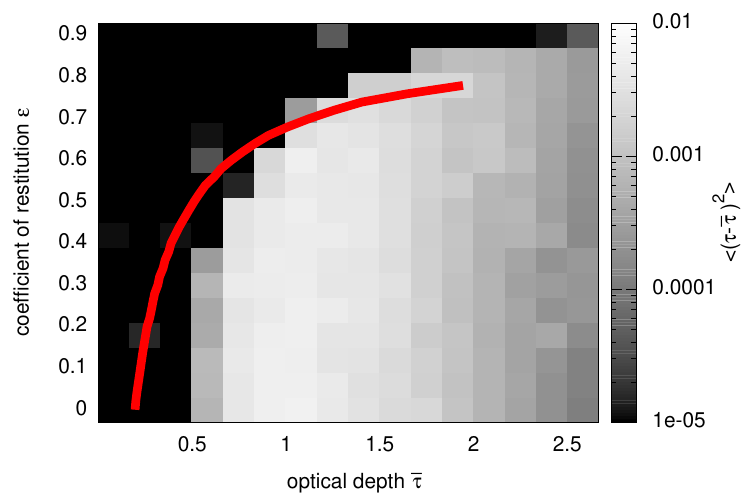}}
 \caption{
  Deviation from the initial optical depth as a function of the initial geometric optical depth and the coefficient of restitution.
  This is a proxy for the activity of the viscous overstability.
  \rev{The red line is an estimate from kinetic theory (see text).}
  \label{fig:array}}
\end{figure}

Next we examine how well the code treats the linear stability of the equilibrium states.
Instead of examining the growth rates of the viscous overstability modes, as in \cite{Schmidtetal2001}, we describe the curve of marginal stability in the full parameter space of $\epsilon$ and $\bar\tau$.
This result extends previous $N$-body work that tested only a handful of cases \citep{SaloSchmidtSpahn2001}, and can directly connect to the predictions of the kinetic modelling \citep{LatterOgilvie2008}.

For given collisional properties, it has been shown that viscous overstability emerges once the optical depth increases above a critical value.
This $\bar\tau$ dependence is actually a proxy for central filling factor $FF_0$, which more directly influences the behaviour of the ring's viscous stress \textbf{(through the collision frequency and close packing effects)}.
As $FF_0$ increases, so does the sensitivity of the stress to density variations, as mediated by the Enskog factor in the kinetic theory \citep{ArakiTremaine1986,LatterOgilvie2008}.
If the stress increases sufficiently fast with density (i.e.\ if $\beta$ is sufficiently large) then instability ensues \citep{SchmitTscharn95,Schmidtetal2001}.

We compute the numerical stability criterion in the following way.
A grid of $\bar\tau$ and $\epsilon$ values is set up and a sequence of simulations run, each corresponding to a unique $(\bar\tau,\epsilon)$.
The simulations use $L_x=1000\,$m and are evolved for up to $10^4$ orbits, which ensures that there is sufficient time for the unstable modes in the box to achieve appreciable amplitudes.
The level of activity in the box is measured by $\langle \left(\tau(t)-\bar\tau\right)^2 \rangle$, where the angle brackets denote a box average.
In order to obtain sufficiently large $FF_0$ for instability at a reasonable $\bar\tau$ we amplify $\Omega_z$ to $3.6~\Omega$.
Doing so mimics the compression of the disk via self-gravitation (see above).
The result of these simulations are plotted in Fig.~\ref{fig:array}, in which we show the magnitude of the activity at the end of the run.
Superimposed on the figure in red is the stability curve calculated from the dense gas kinetic theory \citep{LatterOgilvie2008}.

As any activity corresponds to the onset of viscous overstability, the purely black regions in Fig.~\ref{fig:array} signify stability.
All non-black regions indicate instability, except for a handful of outliers influenced by noise.
Clearly, for given $\epsilon$ less than roughly 0.9 there is a critical $\bar\tau$ above which we have instability, in agreement with previous results.
Larger $\epsilon$ generally requires a larger $\bar\tau$.
This is because the velocity dispersion and disk thickness tend to increase with $\epsilon$, there being less dissipation.
As a consequence, greater filling factors are more difficult to achieve at lower optical depths.
There also appears to be a hard cut-off around $\bar\tau=0.5$ below which instability can never occur.

There is general qualitative agreement with the kinetic marginal curve.
The quantitative discrepancies arise because instability occurs for larger $FF_0$ (above roughly 0.35), the regime in which the kinetic theory is less reliable. 
For larger $\epsilon$ this is especially a problem; \rev{in these less dissipative conditions, $FF_0$ must take very large values ($\approx 0.5$) in order to drive instability.}

In summary, both the equilibrium and linear stability results show that our particle code is behaving in the correct manner, with regards to previous calculations.
This gives us confidence in its capability to describing the nonlinear evolution of the viscous overstability which is the main focus of this paper.

\subsection{Numerical convergence} \label{sec:convergence}
\begin{figure}
		\centering
	\subfigure[Density power spectra for different timesteps $dt$. The unit of the timestep is $2\pi\Omega^{-1}$.]{
		\centering
		\resizebox{0.4\textwidth}{!}{\includegraphics{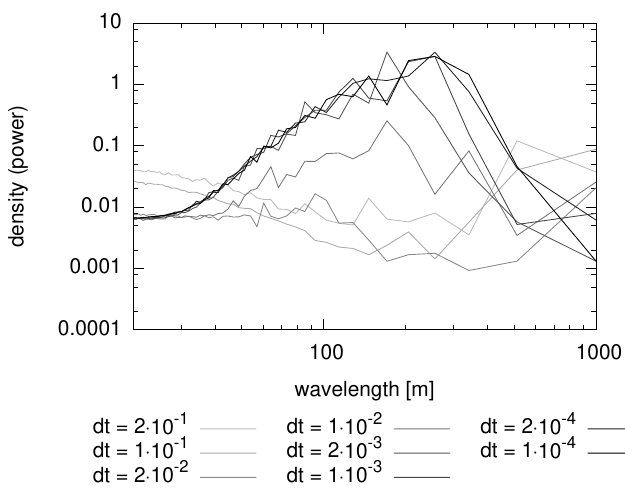}}
		\label{fig:conv:dt}
	}
	\subfigure[Density power spectra for different azimuthal box widths $L_y$.]{
		\centering
		\resizebox{0.4\textwidth}{!}{\includegraphics{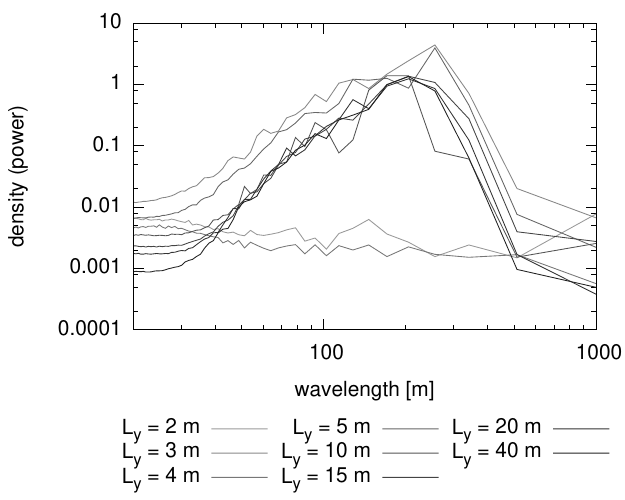}}
		\label{fig:conv:ny}
	}
	\subfigure[Density power spectra for different radial box widths $L_x$.]{
		\centering
		\resizebox{0.4\textwidth}{!}{\includegraphics{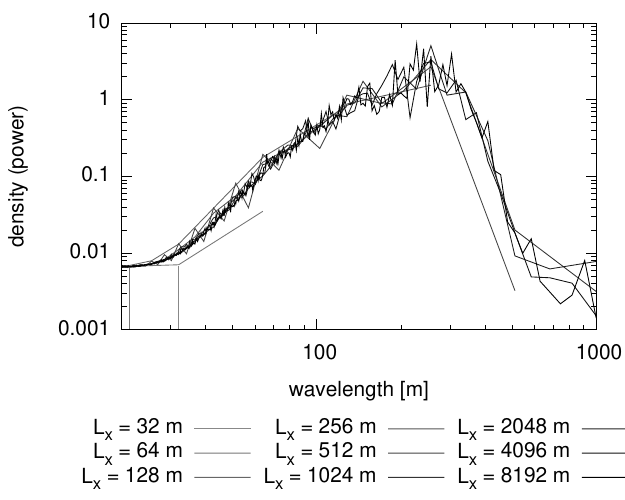}}
		\label{fig:conv:nx}
	}
	\caption{Convergence study. Power spectra of the density for different timesteps and box sizes.}
	\vspace{2cm}
	\label{fig:conv}
\end{figure}
We verify how the results of the simulations depend on the size of the box and the timestep.
These are the only two numerical parameters we need to check.
In these simulations the geometric optical depth is~$\bar\tau=1.64$ and we use a constant coefficient of restitution~$\epsilon=0.5$.
As we have shown in Sect.~\ref{sec:array}, this system will be viscously overstable.

We integrate the system for approximately 320 orbits which gives the overstability enough time to develop.
The power spectra of the radial variations in the density, the average height of a particle above the mid-plane and the average velocity of a particle are then measured 100 times per orbit and averaged.
One of these quantities, the power spectrum of the density, is plotted in Fig.~\ref{fig:conv} for different $dt$, $L_x$ and $L_y$.
The other spectra are qualitatively very similar.

In Fig.~\ref{fig:conv:dt} we fix $L_x=1024\,$m and $L_y=5\,$m and vary the timestep.
One can see that a timestep $dt=2\pi\cdot10^{-3} \Omega^{-1}$ or smaller is sufficient to \rev{obtain convergence in the the overstability power spectrum}.
\rev{Fig.~\ref{fig:conv:ny} shows that, when all the other parameters are fixed, the overstability is suppressed if the azimuthal extent of the box $L_y$ falls below $4$~m.}
\rev{However, when $L_y\geq 10\,$m the spectra are independent of $L_y$.}
\rev{We hence adopt $L_y=15.53$~m for most runs.}
Finally, Fig.~\ref{fig:conv:nx} indicates that a radial box width of $L_x=1024\,$m is sufficient to accurately capture the dominant wavelengths in the nonlinear saturated state.

To summarise, the development of the viscous overstability in these smaller runs is relatively insensitive to the numerical parameters.
Of course this does not tell us whether the \rev{long-term and large-scale} nonlinear behaviour is described adequately, but it nonetheless \rev{gives us confidence} that this is indeed the case.

We should also point out that in simulations with higher optical depth, $\bar\tau\geq2$, we need a larger $L_y$ to avoid crystallisation of particles.
Quasi-crystalline structures are a lower energy state and act as an attractor if the velocity dispersion is too small to randomise the particle motion.
Higher density rings are more prone to this effect because a higher collision rate leads to a cooler and denser ring with strong excluded volume effects and particle correlations.
This is purely an artefact of using equal-sized particles and will not occur in a real ring\footnote{Analogous crystallisation phenomena have been hypothesised as a result of the (poorly constrained) surface adhesion between ring particles by \cite{Tremaine2003}. See also \citet{Perrine2011,Perrine2012}.}.
We could simply allow for a size distribution to avoid this problem  (we present one such simulation in Sect.~\ref{sec:exploration}).
However, this would complicate the analysis and comparison to analytical estimates.

\section{Numerical results}\label{sec:results}
In this section we present our main results on the nonlinear saturation of the viscous overstability in large radial domains and over long times.
We ran simulations varying the two main control parameters $\bar\tau$ and $\epsilon$, but we describe only one set of parameters in detail.
This `fiducial simulation' conveniently exhibits all the main features witnessed in the other runs.
In addition, we tested how different global structures impacted on the long term behaviour.
Specifically, simulations are shown of a spreading ring and a ring with a density gradient.

\subsection{Fiducial simulation}\label{sec:fiducial}
Our illustrative run employs an optical depth of $\bar\tau=1.64$.
We use shearing sheet boundary conditions with an azimuthal and radial width of $L_y=15.53\,$m and $L_x=12.43$~km, respectively.
Initially all particles are randomly distributed.
The simulation is then integrated for 8200~orbits, which corresponds to over 10~Earth-years assuming a distance from Saturn of $a=130000$~km.

\begin{figure}
\centering \resizebox{0.99\columnwidth}{!}{\includegraphics{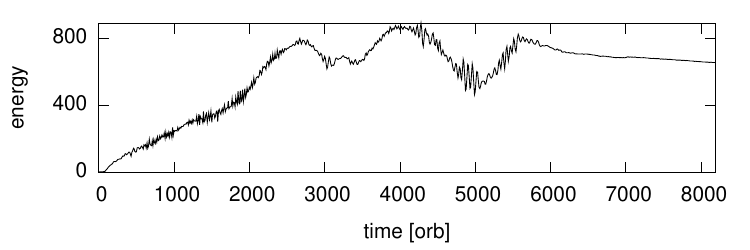}}
\caption{Kinetic energy evolution (only x and y components) in the fiducial simulation. \label{fig:fiducialenergy2}}
\end{figure}
Viscous overstability attains nonlinear amplitudes after a few hundred orbits, and then settles into a disordered state for some 5000 orbits.
After roughly 5300 orbits this state resolves into a single travelling wavetrain of relatively uniform amplitude.
The first disordered stage of the evolution can be arranged into two consecutive phases, (a) `wave turbulence', characterised by \rev{the competition of counterpropagating travelling waves of various wavelengths} and (b) a cellular source/shock state, in which relatively well-defined patches of \rev{travelling waves} are separated by wave defects (shocks and sources).
In Fig.~\ref{fig:fiducialenergy2} we present the total fluctuating kinetic energy of the box as a function of time.
In it the three phases of the evolution can be distinguished.
From roughly 500 orbits to 2000 orbits, there is a linear increase of energy associated with stage (a), the `wave turbulence'.
Then from orbits 2000 to 5300 the energy saturates, while undergoing long-term oscillations.
This corresponds to phase (b), the source/shock state.
After 5300 orbits the system relaxes to a single travelling wavetrain that fills up the entire box and the total energy \rev{decreases towards a constant value}.
This is the end-state of the evolution and is shared by all simulations that used shear-periodic radial boundaries.
We now discuss each of these three phases in more detail.

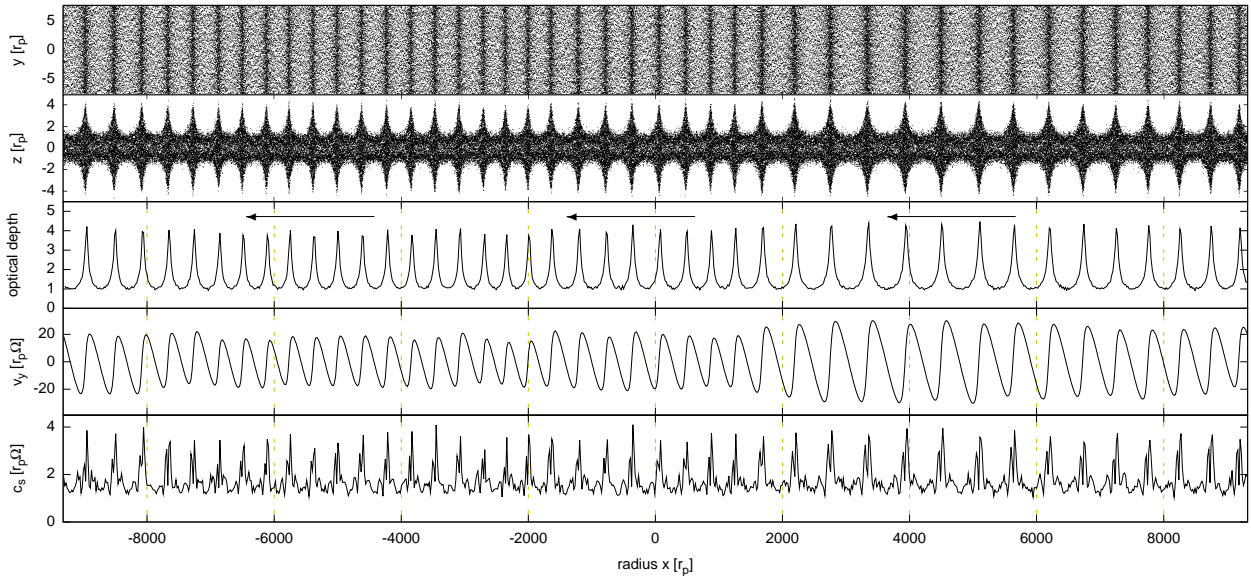
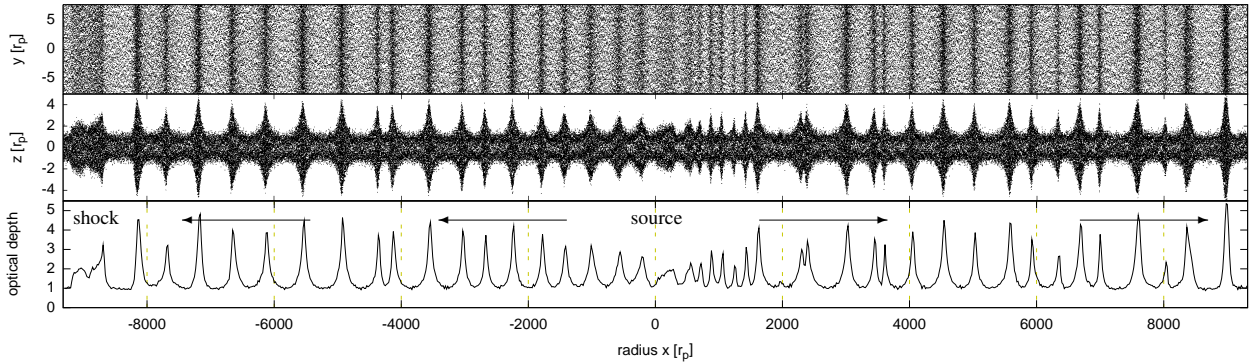
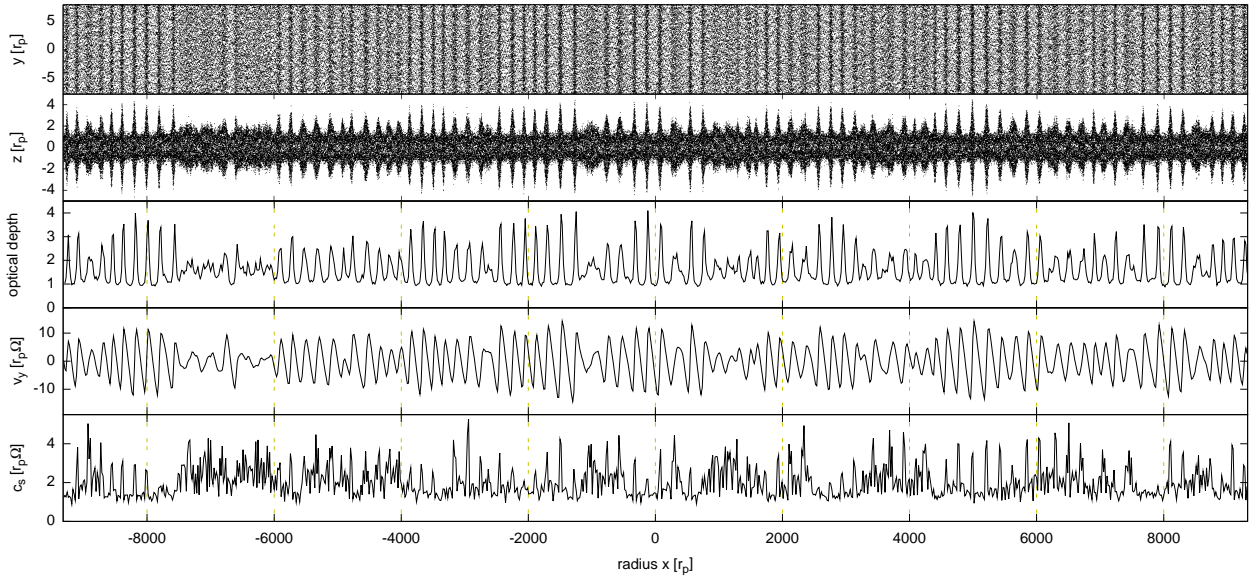
\begin{figure*}
	\subfigure[Fiducial simulation after $t=8000$~orbits.]{
		\resizebox{0.97\textwidth}{!}{
			\begin{overpic}
			{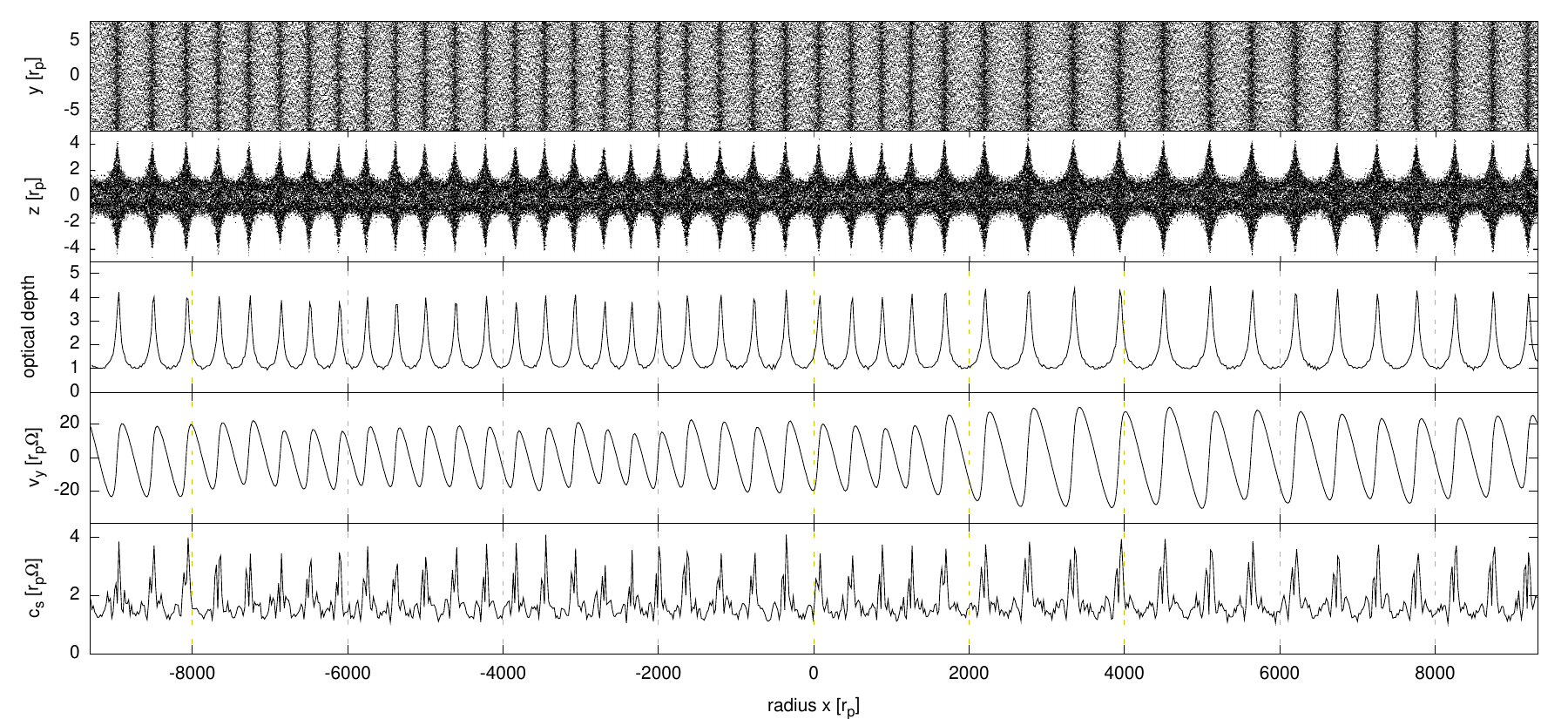}
			 \put(30, 28){\vector(-1, 0){10}}
			 \put(55, 28){\vector(-1, 0){10}}
			 \put(80, 28){\vector(-1, 0){10}}
			\end{overpic}
		}
		\label{fig:fiducial8000}
	}
	\subfigure[Fiducial simulation after $t=4000$~orbits.]{
		\resizebox{0.97\textwidth}{!}{
			\begin{overpic}
			{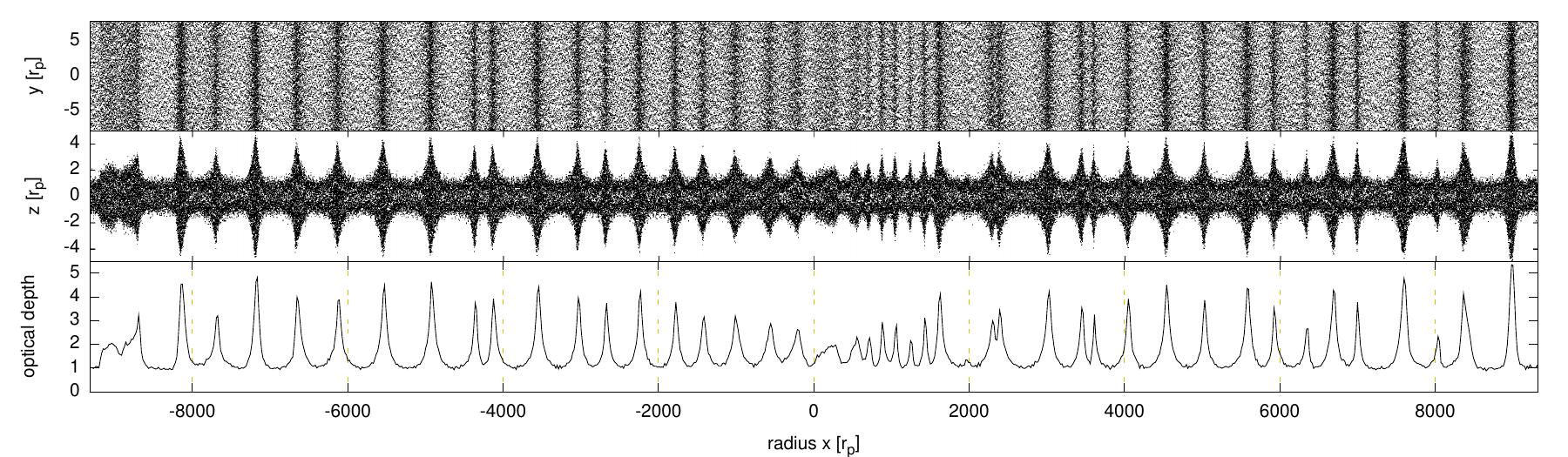}
			 \put(25, 11){\vector(-1, 0){10}}
			 \put(45, 11){\vector(-1, 0){10}}
			 \put(60, 11){\vector(1, 0){10}}
			 \put(85, 11){\vector(1, 0){10}}
			 \put(50, 10.7){source}
			 \put(6.5, 10.7){shock}
			\end{overpic}
		}
		\label{fig:fiducial4000}
	}
	\subfigure[Fiducial simulation after $t=300$~orbits.]{
		\resizebox{0.97\textwidth}{!}{
			\begin{overpic}
			{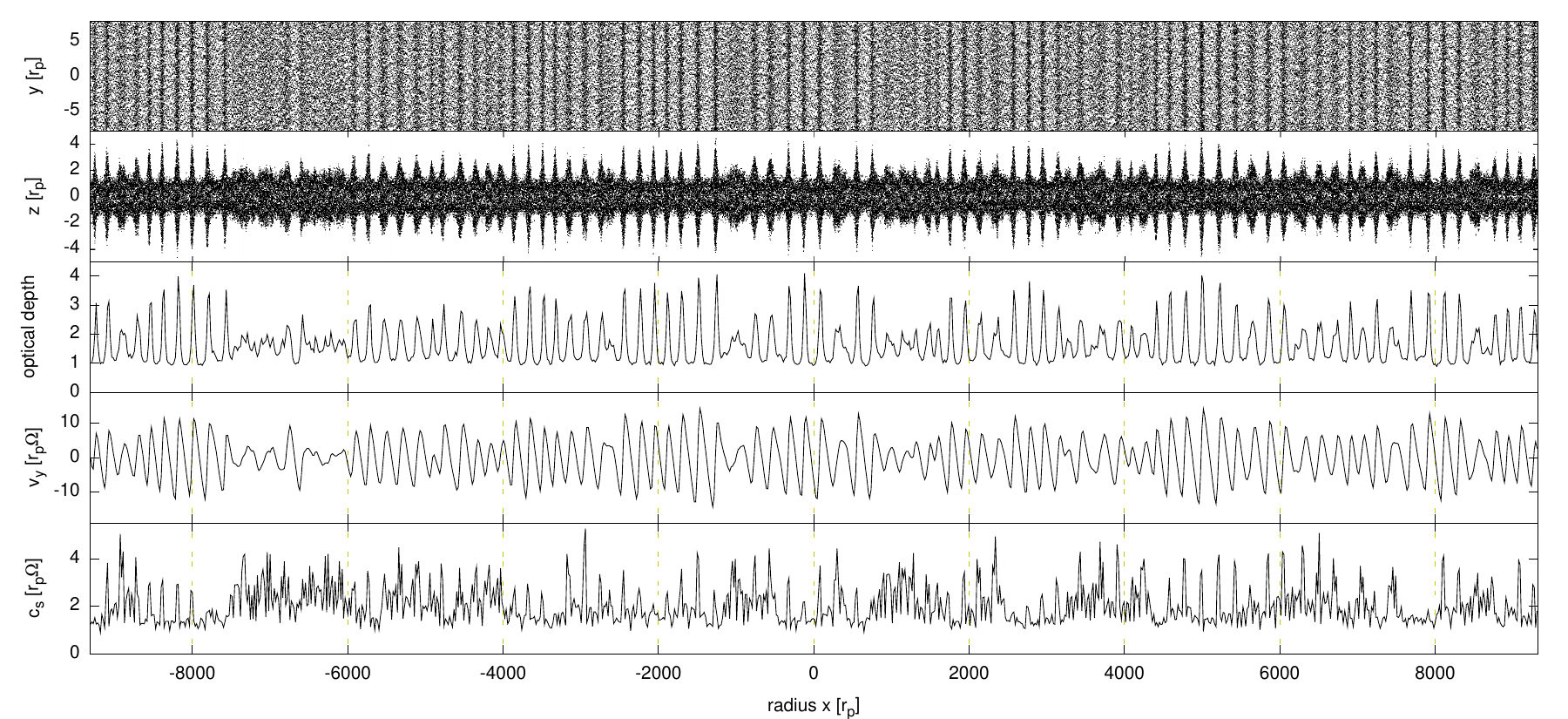}
			\end{overpic}
		}
		\label{fig:fiducial300}
	}
	\caption{Snapshot of the fiducial simulation at different times.}
\end{figure*}
We begin in reverse with the final state, which we plot in Fig.~\ref{fig:fiducial8000}.
This figure shows a snapshot of the particle positions in the $(x,y)$ plane and a $(x,z)$ cross-section, the optical depth $\tau$, the $y$- and $z$-averaged azimuthal velocities (relative to the shear)~$v_y$, and the velocity dispersion $c$ after $t=8000$~orbits.
We find that the final stable nonlinear wavetrain state propagates in a single direction.
The first plot confirms that the nonlinear evolution remains axisymmetric (within the limited azimuthal extent of the box), while the $\tau$ profile exhibits the characteristic cusp-like shape shared by all density waves in thin disks, where particle motions are controlled by epicycles.
The system has yet to completely relax to a wavetrain of uniform amplitude, and displays a large-scale modulation in wavelength.
Ultimately this modulation decays.
The average wavelength exhibited is some $450\,$m.
The velocity dispersion profile reveals spikes in the random motions at wavetrain peaks.
It is in these high density peaks that the orbital energy is primarily thermalised, as the collision frequency (and associated collisional stresses) are maximised there.
The $(x,z)$ cross-section shows these regions of high-density are accompanied by coherent vertical expansions (splashing) via excluded volume effects \citep{Borderies1985}.
This vertical structure is connected to the vertical `breathing mode' exhibited by viscous overstability in gaseous disks \citep{LatterOgilvie2006b}.

The source/shock state is described in Fig.~\ref{fig:fiducial4000}, which is similar to the previous figure but omits the $v_y$ and $c$ profiles.
The $\tau$ profile shows the existence of two patches of density waves of varying wavelength with their direction of propagation indicated by arrows.
The two wavetrains are separated by a source region located at roughly $x=0\,$m, which generates waves, and a shock region located around $x=-9000\,$m, which is where the counterpropagating waves collide and destroy each other.
The shock region straddles the (periodic) radial boundary.
\rev{Clear signatures of these features in a wavelet plot are presented in Appendix~\ref{app:extreme} for a simulation with a much longer domain.}
This configuration of dynamical elements reflects behaviour in the complex Ginzburg-Landau equation, which serves as a good reduced model for the overstable system \citep{AransonKramer2002,LatterOgilvie2009}.
It also reproduces qualitatively the hydrodynamical simulations of \cite{LatterOgilvie2010}, yet there exist interesting discrepancies.
The wavetrains themselves are far more disordered and undergo larger amplitude fluctuations in their amplitude and phase.
In addition, the source and shock regions are broader and less defined.
In particular, the shock region can extend over 1000-1500~m, i.e.\ many characteristic wavelengths of the wavetrain.
This is because colliding wavetrains penetrate each other more successfully in the particle simulations and must undergo multiple `collisions' before they dissipate.
By contrast, the hydro simulations support shocks that are very narrow and well-defined: waves are almost totally destroyed in a single collision.
This difference in the nonlinear interaction between travelling wavetrains is perhaps the key disagreement between the hydrodynamical and $N$-body modelling.

In Figure~\ref{fig:fiducial300} we plot a snapshot of the evolution during the initial \rev{wave turbulent state.}
The dynamics in this phase is complicated, \rev{comprising stretches of counterpropagating travelling waves of different wavelengths interpenetrating each other and giving rise to various nonlinear interactions}, such as beating and standing wave patterns.
\rev{\cite{SchmidtSalo2003} constructed a reduced model of this nonlinear competition for two counterpropagating travelling waves with the same (marginal) wavelength.
For realistic parameters, they show that ultimately one of the two travelling waves suppresses the other, and that no standing wave solution is stable. 
Our simulations describe this competition in its full generality, including wave interactions between all scales above 50~m. 
In agreement with \cite{SchmidtSalo2003} we do not find stable travelling wave solutions.}
As is clear by comparison with Fig.~\ref{fig:fiducial4000}, characteristic wavelengths are much shorter in this evolutionary stage, closer to the wavelength of the most unstable mode ($\sim 100\,$m), and they steadily increase with time \citep{SchmitTscharn1999}.
This, in fact, drives the increase in kinetic energy witnessed in Fig.~\ref{fig:fiducialenergy2}, as the wave amplitude \rev{is proportional to wavelength \citep{SchmidtSalo2003,LatterOgilvie2009}}.
This stage of the evolution bears similarities with the one-dimensional `wave turbulence' that characterises certain plasmas, nonlinear optics, and other physical phenomena \citep[for e.g.][]{Nazarenko2011}. 
Notably, the overstable ring shares with those systems an inverse cascade of energy to longer lengths and a general movement towards greater nonlinearity and coherent structure.

\begin{figure}
\centering \resizebox{0.99\columnwidth}{!}{\includegraphics{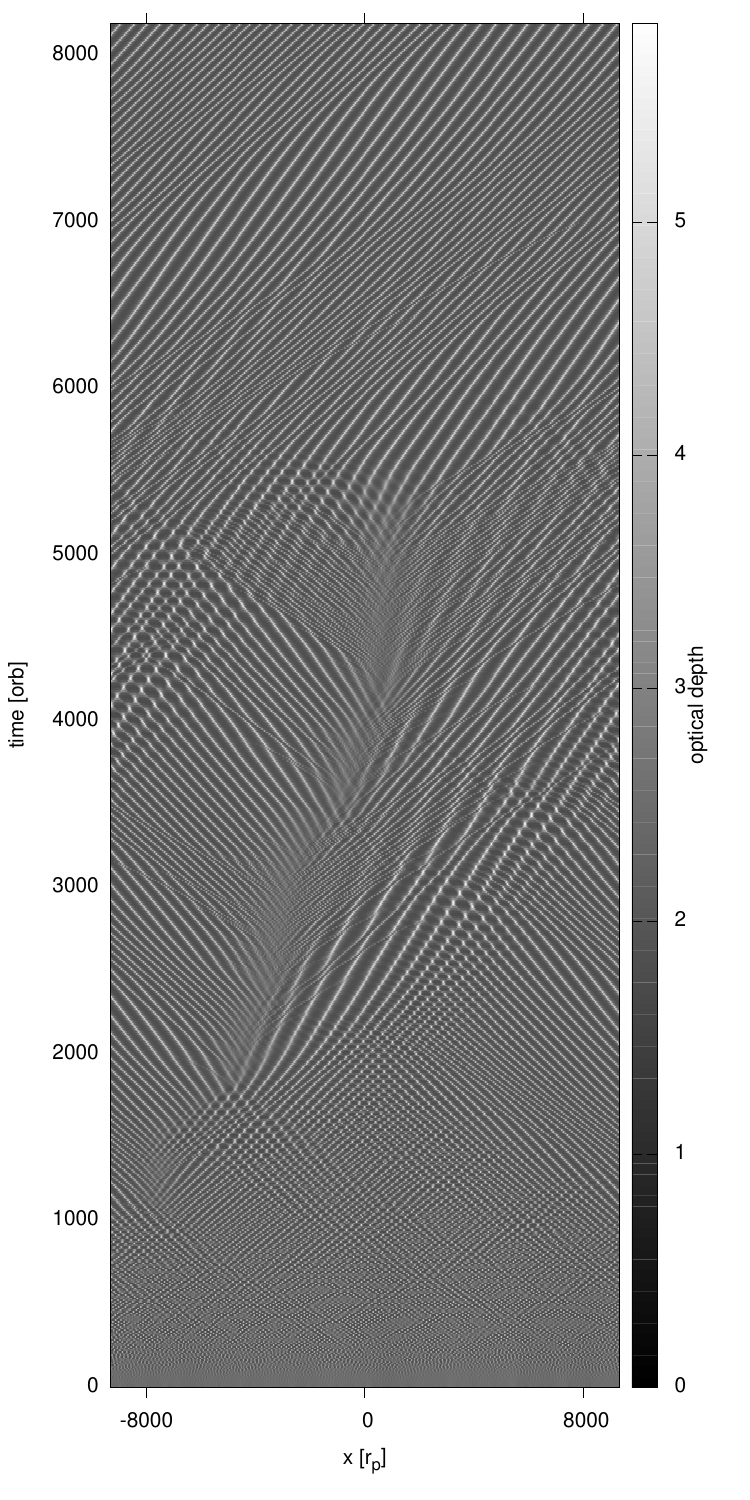}}
\caption{Stroboscopic space-time plot of the fiducial simulation. \label{fig:fiducialspacetime}}
\end{figure}
The stages of the dynamical evolution can be summarised in a stroboscopic space-time diagram, Fig.~\ref{fig:fiducialspacetime}, in which the optical depth is plotted as a function of $x$ and $t$, but only at every 10th orbit.
This screens out the fast time associated with the wave phase speed yet keeps the sense of its and the group velocity's direction, while also indicating important features such as the sources and shocks.
The first clear signal of a source region emerges at around $t=1800\,$orbits near $x=-5000\,$m.
It can be distinguished by the light diagonal rays propagating away on either side.
These rays represent the slower group velocity of the waves (not the very rapid phase velocity $\sim 100\,$m per orbit).
The shock region is less easy to pick out.
At $t=2000\,$orbits it extends from $x=-1000\,$m to $x=1000\,$m.
As illustrated by hydrodynamical simulations, source and shock regions migrate around the domain on a slow time-scale.
Eventually they collide and annihilate each other.
This occurs near $t=5300\,$orbits.
After this point we have a single wavetrain propagating in one direction (all the rays point in the same direction).
Another interesting feature is that the wavelength of the waves generated by the source vary with time and can be different depending on which side of the source they originate.
The time-variable behaviour was noted in the hydrodynamical simulations but was less prominent; the \rev{wavelength} asymmetry about a source\rev{, on the other hand,} appears a novel feature of the $N$-body simulations.

In all the runs undertaken with shearing periodic boundaries the long term saturation of overstability is a single travelling wavetrain of uniform amplitude.
However, this final state is to some some extent an artefact of the shearing box's translational symmetry in radius.
As far as the collective axisymmetric dynamics are concerned, there is no preferred radius.
A real ring, on the other hand, exhibits radial structure in particle size, composition, and number density on intermediate to long scales.
Moreover, resonances and wakes create spatially fixed perturbations.
These variations should prohibit the transition to a smooth and coherent state with one wave-train.
It is therefore likely that the real rings will exhibit the dynamics exhibited by simulations on \emph{intermediate times}, before the influence of the boundary conditions dominate --- in other words, before, the box realises that it is periodic.
Viscous overstability in the real rings will probably manifest as either (a) \rev{the wave turbulent state} or (b) the source/shock state.
Perhaps different regions in the rings support one or the other state, giving rise to differing observations on intermediate scales
$1-10$~km.
As revealed by the Cassini cameras, there exists a rich and irregular set of radial variations on these scales \citep[Figures 5A and 5F in ][]{Porco2005}.
In Section~\ref{sec:structure} we trial other global set-ups and boundary conditions in order to break the radial translational symmetry and to assess how the overstability saturates in the presence of simple global structures.

We ran one additional simulation with the same parameters as the fiducial simulation but with an extremely large radial domain, spanning over 55\,000~particle radii (55~km).
More details of the larger simulation, as well as a stroboscopic space-time diagram \rev{and wavelet analysis}, are presented in Appendix~\ref{app:extreme}.
As is discussed there, precisely the same phenomena appears on similar scales in the large box.
\rev{The results} show the box independence of our results.
\rev{They} also suggests that the longest scales generated by viscous overstability do not exceed 5-10~km.

\subsection{Exploration of different parameters}\label{sec:exploration}
In order to test the robustness of \rev{our} results, we conducted simulations with different $\epsilon$ and $\bar\tau$.
In particular, we used both $\epsilon=0$ (\rev{complete damping of the normal component of the relative velocity}) and the velocity dependent \cite{Bridges1984} collision law while keeping $\tau=1.64$.
Neither resulted in nonlinear evolutions significantly different to that of the fiducial run.
This makes us confident that, for sufficiently inelastic collisions, the precise treatment of the coefficient of restitution is unimportant and that a value of $\epsilon=0.5$ yields representative results.
It is likely that larger $\epsilon$ may lead to different nonlinear behaviour, as the ring will generally be `hotter' and less dense.
However, experimental and theoretical evidence points to very dissipative collisions \citep{Porco2008, Schmidt2009}, so this hot regime is probably irrelevant to conditions in Saturn's rings.

We undertook additional large-scale simulations with two different $\bar\tau$, equal to 1 and 2.
In each case $\epsilon=0.5$.
In neither case, was the general scheme of the nonlinear evolution different to the fiducial case.
It should be noted, however, that the lower $\bar\tau$ run exhibited \rev{fewer violent fluctuations and its wavetrains were more ordered} during the source/shock phase.
What these simulations also revealed was that the characteristic wavelength of the dynamics (in which the power was concentrated) positively correlated with optical depth.
Lower $\bar\tau$ runs yield shorter wavelength behaviour, while higher $\bar\tau$ gives a longer characteristic wavelength.
This correlation was also noted indirectly in \cite{LatterOgilvie2009} in their Table 3, which shows that the wavelength of the first stable wavetrain increased with $\bar\tau$ (as mediated by the dimensionless parameter $\beta$).
It is likely that in the disordered nonlinear state, this first stable wavelength works as an attractor and fixes the characteristic wavelength on sufficiently long times.

\begin{figure}
\centering \resizebox{0.99\columnwidth}{!}{\includegraphics[trim=0.5cm 0.4cm 0.0cm 0.9cm]{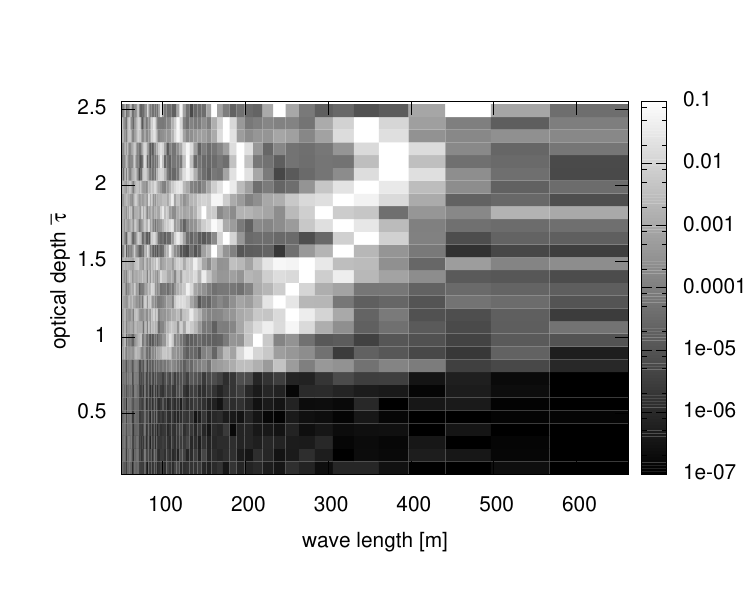}}
\caption{Power (arbitrary units) as a function of wavelength (horizontal axis) and the optical depth $\bar\tau$ (vertical axis). Each row corresponds to one simulation. \label{fig:wavelength}}
\end{figure}
To develop this idea we subsequently ran a sequence of smaller scale simulations for different $\bar\tau$, in which $L_x=4000\,$m.
Each simulation was run for 5000 orbits.
After this length of time the system resolved into a single stable wavetrain, because of the smaller box size.
Figure~\ref{fig:wavelength} summarises these results by plotting the power in each wavelength for each $\bar\tau$.
As is clear, the wavelength of most power increases linearly with optical depth, beginning at $210\,$m at $\bar\tau=1$ and rising to roughly $400\,$m at $\bar\tau=2$.
Note that the viscous overstability is not active for $\bar\tau<0.75$, consistent with the results in Sect.~\ref{sec:array}.
A correlation of wavelength with $\tau$ may be evident in recent \emph{Cassini} observations (M.~Hedman, private communication).

\begin{figure*}
\centering \resizebox{0.96\textwidth}{!}{\includegraphics{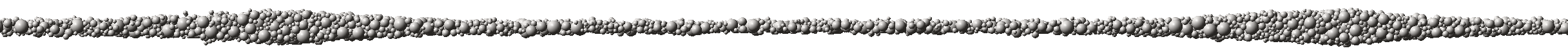}}
\caption{3D rendering that shows a small part of the $(x,z)$ plane in the simulation using a particle size distribution. \label{fig:sizedistr}}
\end{figure*}
All simulations presented so far use a single size of ring particles. 
We also ran an additional simulation using a size distribution.
The distribution \rev{follows} $dN/dr_p\propto r_p^{-3}$ and $r_p$ ranges from $0.5$~m to $2$~m, motivated by current estimates of the typical size distribution in Saturn's rings \citep{Cuzzi2009}.
The system is viscously overstable and the non-linear behaviour is indistinguishable from the fiducial simulation.
As an illustration, we present a 3D rendering of a small part of the $(x,z)$ plane in Fig.~\ref{fig:sizedistr}.
The plot shows approximately two wavelengths.
Clearly visible is the vertical splashing of particles.
These simulations make us confident that our results are general and are not an artefact of the single sized ring particles. 
There might be quantitative differences that depend on the precise nature of the size distribution. 
However, our results indicate that these effects will be small and we therefore do not investigate them any further in this paper. 

\subsection{Exploration of different global structures}\label{sec:structure}
In this subsection we extend the shearing box model to account for different global structures: a large spreading ring and a ring with a radial density gradient.
We then check how these influence the long term saturation of the overstability.
The large extent of our simulations (covering scales separated by almost 5 orders of magnitude --- $1\,$m to $50\,$km) allows a convincing exploration of global structure in direct $N$-body simulations.

\subsubsection{Open boundary conditions}
\begin{figure*}
	\subfigure[Stroboscopic space-time diagram of a spreading ring with open boundary conditions.]{
		 \resizebox{0.93\columnwidth}{!}{\includegraphics[trim=0 0 0.6cm 0]{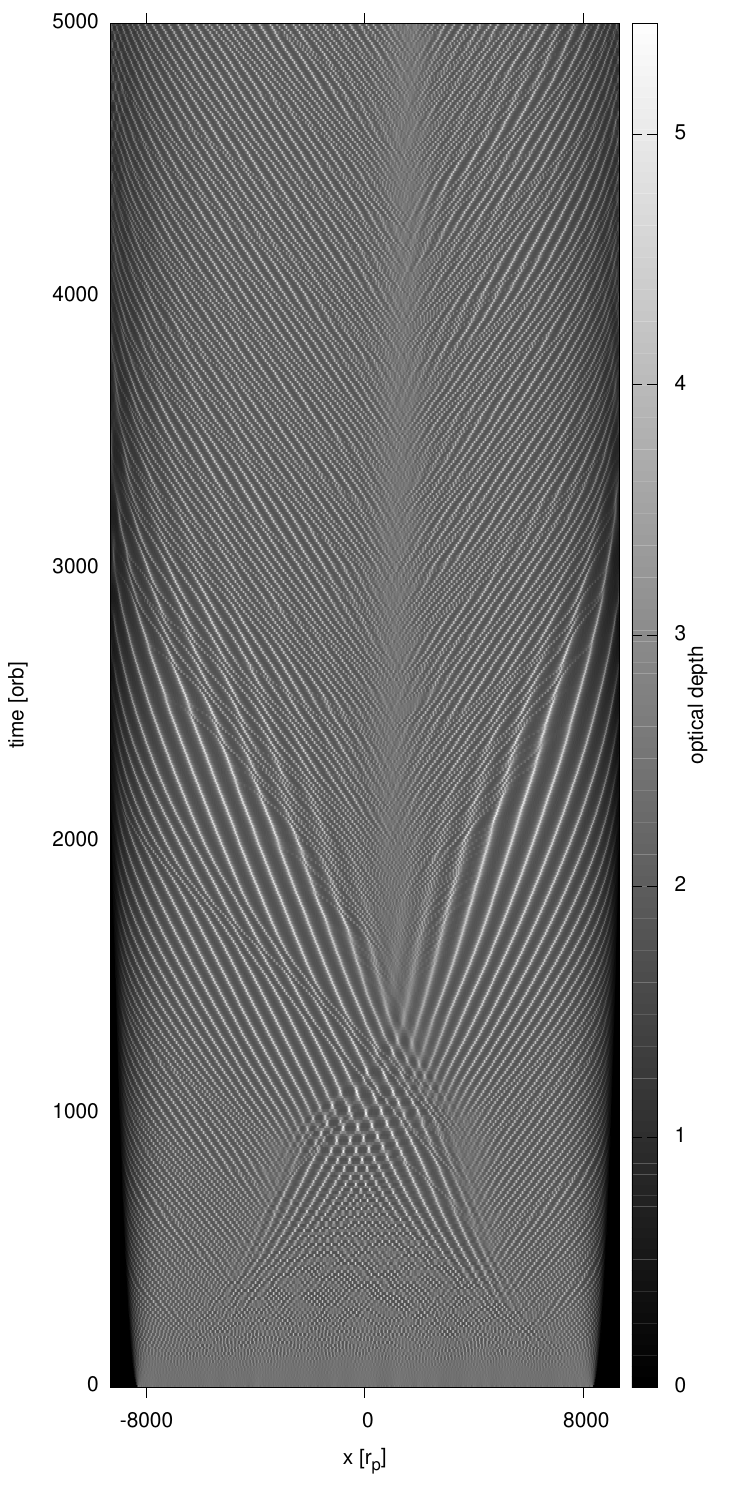}}
		\label{fig:spreading}
	}
	\hspace{0.02\textwidth}
	\subfigure[Stroboscopic space-time diagram of a ring with an initial surface density gradient.]{
		 \resizebox{0.93\columnwidth}{!}{\includegraphics[trim=0 0 0.6cm 0]{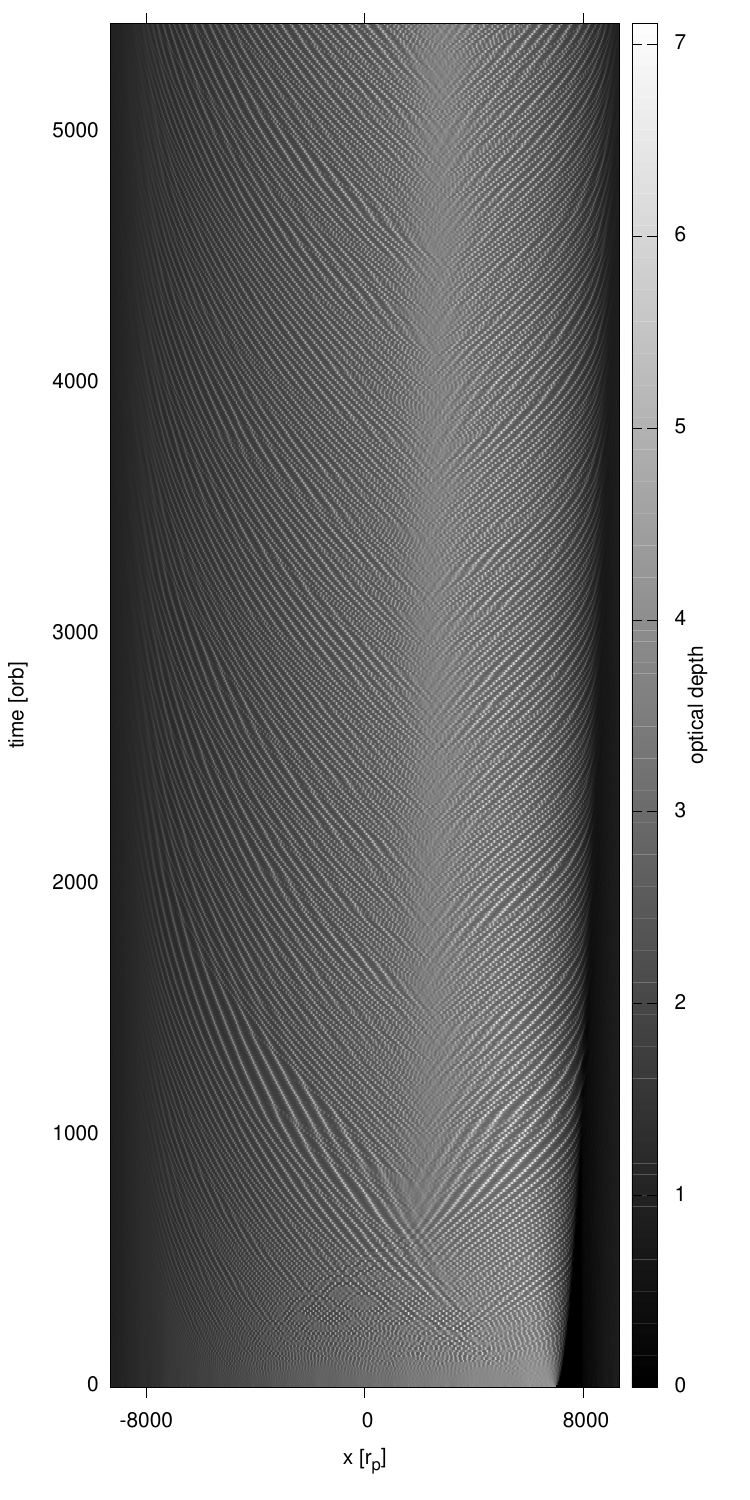}}
		 \label{fig:gradient}
	}
	\caption{Stroboscopic space-time diagram of different global structures.}
\end{figure*}
By open boundary conditions we mean a ring that has a finite radial width less than the radial box size $L_x$.
Thus the ring can spread viscously.
The boundaries are, in fact, not really open: if particles exit the box on one side, they enter the box on the other side.
However, we initially place the box boundaries far from the disk edge ($2000\,$m) and stop the simulation when the ring has spread to the point that particles cross the boundary.

We present a space-time plot of such a simulation in Figure~\ref{fig:spreading}.
One can see that the ring edges are indeed spreading viscously (i.e. $\sim\sqrt{t}$).
The development of the overstability in the middle of the ring \rev{remains} unchanged by these boundary conditions: the instability is local on early and intermediate times.
In fact, waves, once they reach the ring edge, are absorbed and thus lost to the system; consequently, the spreading edge behaves like an open boundary or the buffered boundaries employed in \cite{LatterOgilvie2010}.
Near $t=500\,$orbits two sources have manifested; one at $x=-4000\,$m and the other at $x=4000\,$m.
After 1200~orbits, after a complicated interaction, the intervening shock region annihilates \rev{one of the sources leaving only a single source near $x=0$.}
This remaining source stays at this locations for the rest of the simulation, as in the hydrodynamical simulations of buffered boundaries.

This persistence of a single source is the key difference to the \rev{earlier simulations of Section 4.1, which used period boundary conditions.}
Though static, the source still exhibits interesting dynamics.
The wavelength of the waves it generates can vary, as at $t=1400\,$orbits when the wavelength drops abruptly to a much shorter value.
The cause of this behaviour is unclear, as is its connection to the preferred wavelength of the system, which we have argued earlier is set by the first linearly stable wavetrain.

We believe that such a setup might be the most realistic simulation of the long term behaviour of the overstability.
No spurious effect from the periodicity can dominate the nonlinear dynamics.
Note that in this setup it is the shock/source state that is favoured; not the earlier disordered wave turbulence.
It is then most likely that it is the source/shock state that dominates in the real rings.

\subsubsection{Surface density gradients}
We know that the surface density in Saturn's rings is not constant.
For example, density waves can create \rev{large-scale} enhancements and deficits in the local optical depth that might affect the overstability in a realistic ring.

To approximately account for \rev{general large-scale density variations}, we ran simulations with an initial linear surface density gradient varying from 0.04 to 3.24.
At the right edge, we drop the optical depth to zero abruptly.
Therefore the right edge is again expected to viscously spread (see previous section).
\rev{The left side (the ramp) will also spread viscously, but on a much longer timescale as the gradient is much smaller.}
We present a space-time plot of our run in Figure~\ref{fig:gradient}.

As in earlier runs, the overstability reaches nonlinear amplitudes within a few hundred orbits.
Note that the ring is initially stable where $x<-7000\,$m because the optical depth is too low.
Similarly to the open boundary case presented \rev{in Section 4.3.1}, one source region eventually wins and is maintained for the rest of the simulation.
The waves that it generates march leftwards into the lower optical depth region and steadily decrease in amplitude and wavelength, an interesting effect analogous to that shown in Section~\ref{sec:exploration}.
Note that the \rev{wave rays} seem to bend upwards on the left side of the plot.
This is because the wavelength/group-velocity depends on optical depth.

\subsection{Synthetic observations}\label{sec:observations}
\begin{figure*}
 \centering \resizebox{0.99\textwidth}{!}{\includegraphics{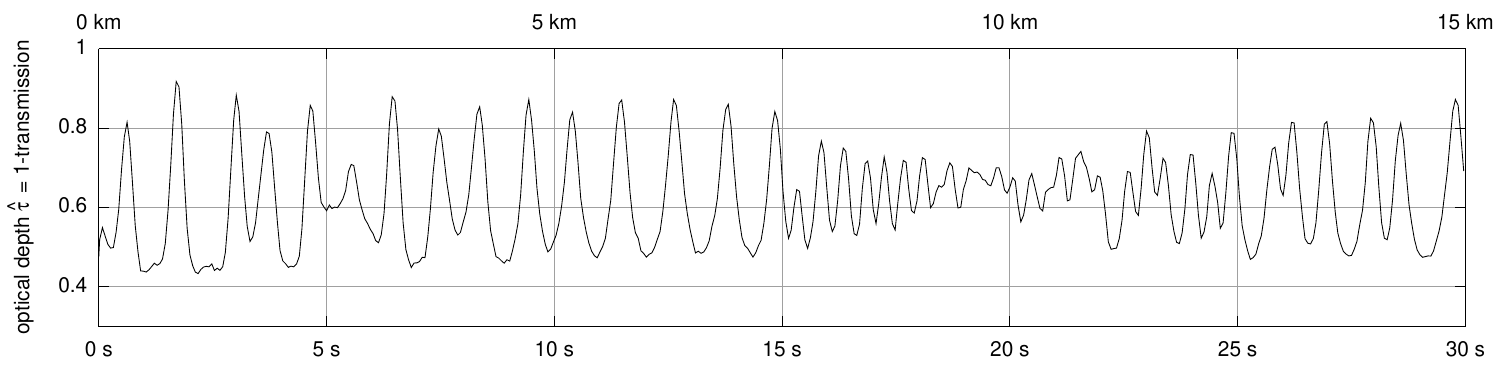}}
 \caption{Synthetic stellar occultation observation as it may be seen by \textit{Cassini} using the fiducial run after
  $t=4000$~orbits.
  \label{fig:observation}}
\end{figure*}
Although this paper is focused on the numerical and theoretical study of the viscous overstability, we present one result from a synthetic stellar occultation observation.
In such an observation, the spacecraft looks through the rings onto a background star.
The star has a finite size and because both Saturn and the spacecraft move, the star effectively moves across the ring.
By measuring the photometric optical depth $\hat\tau$ or the transmission (the fraction of the light not blocked by ring particles) one can estimate the local optical depth.

Let us assume an effective stellar radius of
$r_\star=70\,$m, an integration time of the photometer of $\Delta t=60\,$ms, a sampling length of $\Delta l=302\,$m (this could also be expressed in terms of a relative velocity) and a path of the star which has a projected angle of $\phi=5.7^\circ$ relative to the azimuthal direction.
These parameters were chosen to allow a comparison with observations that can be made with the Cassini spacecraft \citep[M.~Hedman, private communication]{Colwelletal2007}.

The result of applying this convolution is shown in Figure~\ref{fig:observation}.
We use a snapshot of the fiducial simulation after $t=4000\,$orbits and plot $\hat \tau$ ($1-$~transmission) as a function of the time of measurement and radial distance (relative to the starting point).
The convolution (integration) using the above parameters makes it different from the optical depth plotted in Figures~\ref{fig:fiducial300} and~\ref{fig:fiducial8000}.
Sharp features are gone, but one can still easily see the radial structure created by the viscous overstability.
Note that the amplitudes are of order $\sim 0.4$ \rev{from crest to trough, as opposed to $\sim 3$ in the $\tau$ profiles in Figure 5.}
The snapshot was chosen so that not all wave-trains have merged and a source is present just right of the centre in the plot.

Early data from the Cassini spacecraft suggests that structures like these are indeed observed.
If confirmed, a more detailed and systematic comparison is needed to fully uncover the underlying ring structure.
This work should be seen as a proof of concept.
We make the code to produce both the simulations and the synthetic observations freely available (see discussion section) so as to jump-start future investigations.

\section{Discussion}\label{sec:discussion}

We have undertaken the first local $N$-body simulations of the viscous overstability in planetary rings over the large spatial and temporal scales necessary to fully describe its nonlinear saturation.
With radial domains extending to 50 km and timespans to $10^4$ orbits, our particle simulations of overstability are the largest yet attempted and can begin to compete with one-dimensional hydrodynamical calculations. 
These great scales are achieved by \rev{new numerical methods implemented in the \reb code and by severely narrowing the azimuthal box size, a permitted saving because the collective dynamics is essentially axisymmetric.}

The simulated dynamics exhibit a wealth of interesting features on different lengths and times ranging over the dimensions of our simulations.
In particular, these features appear to correspond to both the periodic microstructure ($\sim$200 m) \citep{Colwelletal2007,Thomson2007} and \rev{possibly the} irregular intermediate variations ($\sim$1-10 km) observed in Saturn's A and B-rings \citep{Porco2005}.
They can also can be directly compared to large-scale hydrodynamical simulations and theory.
We may then differentiate robust dynamical behaviour common to both and which we then expect to prevail in the real rings.

The most pervasive dynamical features in our numerical work are axisymmetric nonlinear density wavetrains, propagating either inward or outward. 
Their wavelength varies between roughly 200 m and 400 m, with the longer waves supported by disks with higher optical depth $\bar\tau$.
We believe that the nonlinear wavetrains can be unambiguously connected to the observed periodic microstructure, though we recognise that our simulated waves are slightly longer than those observed with \emph{Cassini}'s occultation experiments
\rev{\citep{Colwelletal2007,Thomson2007}.}
In addition, we find that \rev{stretches of travelling overstable waves interact in complicated ways.}
Earlier in the evolution, superpositions of multiple wavetrains form a disordered field of nonlinear standing modes and beating patterns (`wave turbulence').
On intermediate times, however, these resolve into a cleaner structure comprising radial patches of waves separated by relatively well defined `boundaries', i.e. source and shock regions (wave defects). 
Shock regions are where counterpropagating wavetrains collide, sources where they are generated and move apart from each other. 
Each wavetrain patch extends over many kilometres, whereas the wave defects are shorter --- typically about 1 km in radial size.
It is possible that this longer structure, associated with the sources, shocks, and wavetrain patches, corresponds to irregular intermediate scale variations observed by the \emph{Cassini} cameras \rev{\citep[Figures 5A and 5F in ][]{Porco2005}.}
\rev{We caution, however, that this connection must remain speculative, not least because observed periodic microstructure (a clear signature of overstability) does not always coincide with the observed intermediate structure (J.\ Schmidt, private communication).}
Finally, we do not find variations generated on scales longer than 10~km, nor do our longest features resemble the observations of 100~km scale features in the B-ring. 
We conclude that these are \emph{not} generated by viscous overstability.
The origin of 100~km structure may lie in ballistic transport instability, electromagnetic instability, or both \citep{Durisen1995,LatterOgilvie2012,GoertzMorfill1988}.

All our simulations with shear periodic boundaries ultimately relax to a similar long-term state: a single wavelength wavetrain propagating in one direction. 
This state should correspond to the shortest stable wavetrain permitted in the box.
However, a single uniform wavetrain solution is not a realistic outcome in the real rings, and is surely an artefact of the periodic boundaries and the neglect of any radial structure, i.e. the assumption of translational symmetry in radius.
We conducted additional simulations that broke this symmetry. 
First we examined a spreading ring, whose edges were far from the box's radial boundaries. 
This set-up approximates outflow boundary conditions or buffered boundaries because waves that impact on the ring edges do not reflect and are lost.
We also simulated a disk with a radial density gradient. 
In both cases the final single wavetrain solution was averted, with the system settling down to a state characterised by a single source region radiating density waves \rev{both inwards and outwards.}
These simulations suggest that the overstability saturates in the real rings through an arrangement of couterpropagating wavetrains interspersed by sources and shocks, embedded in the underlying disk structure.
However, there may still exist ring regions stuck in wave turbulence, exhibiting different photometric properties.

Overall the results from the particle code are in good qualitative agreement with comparable hydrodynamical theory and simulations \citep{SchmidtSalo2003,LatterOgilvie2009,LatterOgilvie2010}.
The dominance of the traveling wavetrains, the emergence of sources and shocks, and the long-term single-wavetrain states are dynamical behaviours shared by both approaches.
Nevertheless, interesting discrepancies exist. 
Importantly, particle simulations of colliding wavetrains show greater interpenetration of the waves than in hydrodynamics where, in general, colliding waves are destroyed immediately; \rev{the wave-wave nonlinearities are different in the granular flow, evidently.}
This leads to much broader shock regions and a longer and more complicated early stage of travelling wave competition. 
Source regions are also wider, while the low level chaotic variability in amplitude and phase witnessed in hydro is much exacerbated in the $N$-body dynamics.

Lastly, we performed a brief set of synthetic occultation observations of our numerical output. 
Direct comparison of such results with real occultation data (such as from \emph{Cassini} UVIS) could in principle allow  \rev{researchers to constrain} a range of particle properties, such as size, density, and $\epsilon$. 
One extension of this work could involve synthetic observations of reflected light, such as those calculated by \cite{SaloKarjalainen2003}, \cite{SaloKarjalainenFrench2004}, and \cite{SaloFrench2010} who successfully explain observations of azimuthal brightness asymmetry, the opposition
effect, and the tilt effect. Applying these techniques to the simulated intermediate scale structure would directly connect overstability to the observations on 1-10 km from Cassini Imaging Science \citep{Porco2005}.

Future work will also assess the influence of self-gravity, \rev{a physical effect that should alter some of our conclusions.}
First, how does it change the axisymmetric dynamics, such as the properties of the wavetrains and the larger-scale structure that play upon them?
Second, how does it change the non-axisymmetric dynamics via self-gravity wakes?
On larger domains and over longer times, can the wakes suppress overstability or vice-versa? 
What is the nature of their interference?
These last questions also lead to issues concerning the longitudinal coherence of the collective axisymmetric dynamics.
Over what azimuthal extent do the density waves (and other features) remain organised?
What is the characteristic coherence length and what determines it?
Obviously, to answer these questions we must move on from the small $L_y$ regime employed here.

We make the entire source code publicly available under the GPL license.
We want to encourage other scientists to both reproduce and expand upon our results.
The source files are provided as a git repository and hosted on github at \url{http://github.com/hannorein/rebound/tree/overstability}.
See \cite{ReinLiu2012} on further informations about the \reb code.
Furthermore, the authors are happy to answer any questions that might arise with the installation.

\section*{Acknowledgments}
\rev{The authors would like to thank Juergen Schmidt for his thorough review. 
His many comments improved the quality of the final manuscript.}
Hanno Rein was supported by the Institute for Advanced Study and the NFS grant AST-0807444.
Henrik Latter acknowledges financial support from STFC grant ST/G002584/1.
We thank Scott Tremaine, Matthew Hedman, Matthew Tiscareno, Joe Burns, Jim Jenkins, Eiichiro Kokubo, and Gordon Ogilvie for helpful comments at various stages of this project.
All plots were made with the gnuplot package.
A large fraction of this work has been conducted at Small World Coffee in Princeton.

\bibliographystyle{mn2e} \bibliography{full}

\appendix

\section{Extremely large radial domain} \label{app:extreme}
\begin{figure*}
 \centering \resizebox{0.99\textwidth}{!}{\includegraphics[trim=0 0 0.6cm 0]{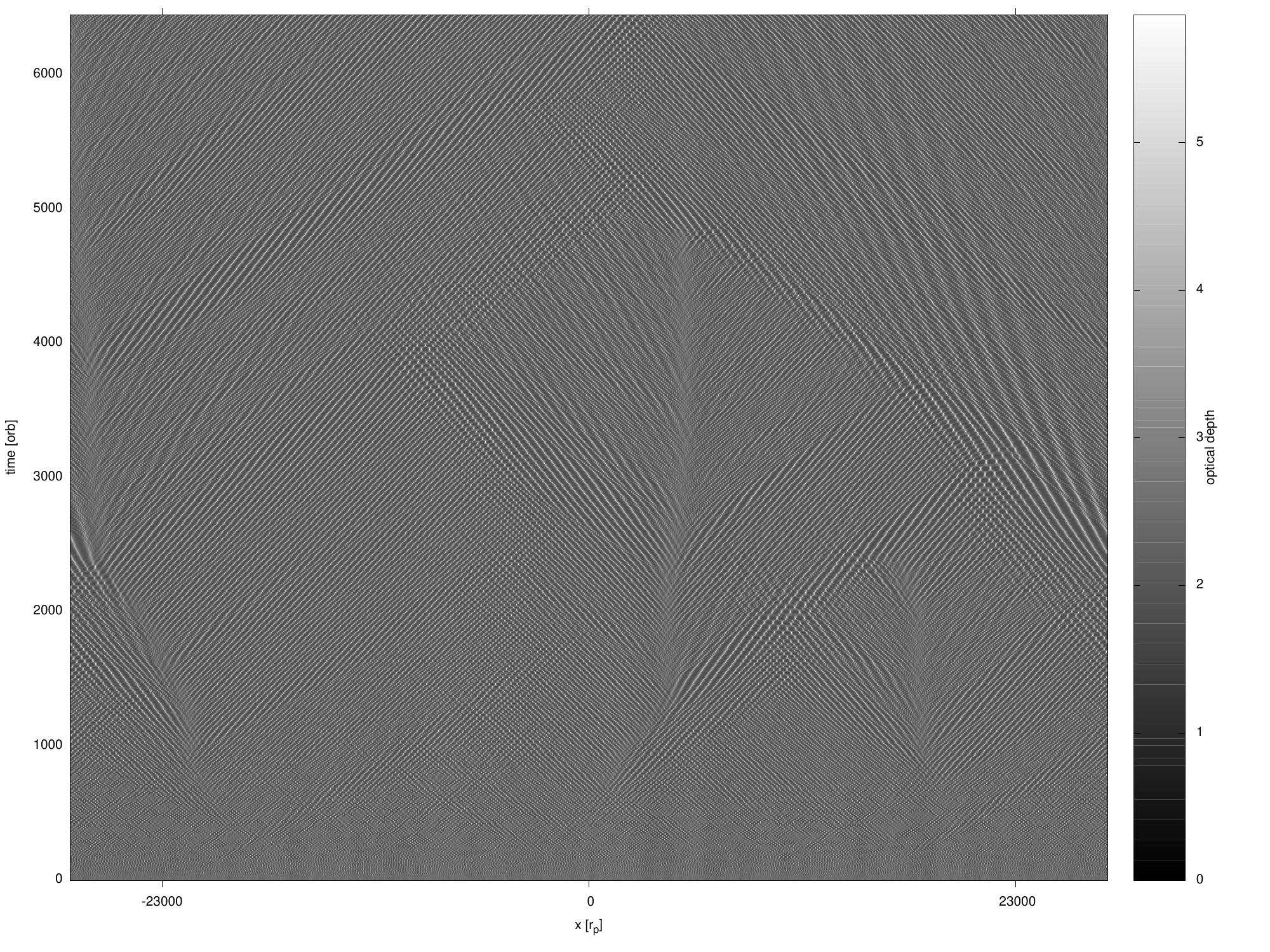}}
 \caption{Simulation with an extremely large radial domain. \label{fig:verylarge}}
\end{figure*}
\begin{figure*}
\centering \begin{overpic}[scale=0.85]{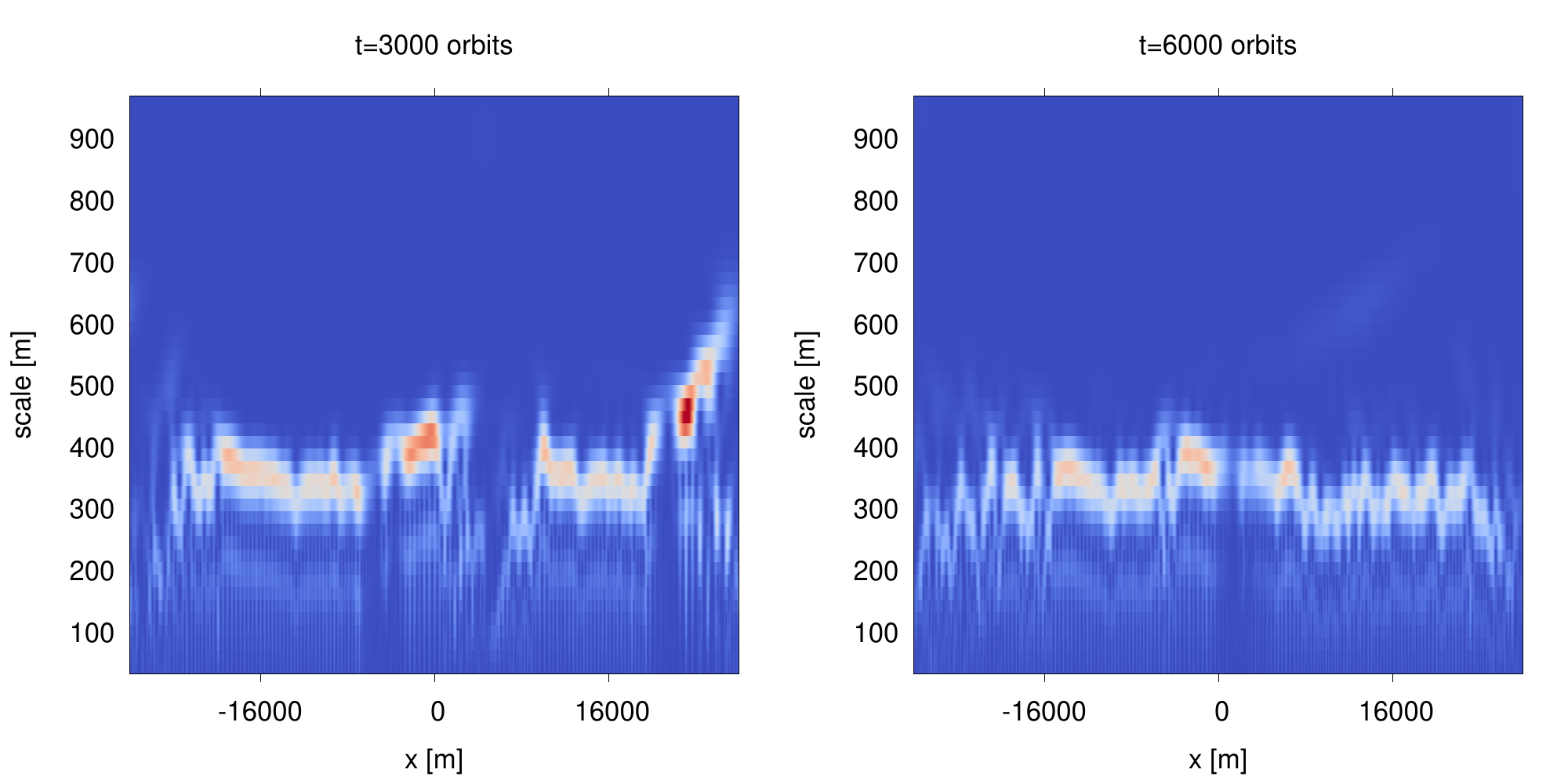}
\put(24,30){\color{white}\begin{rotate}{90}shock region\end{rotate}}
\put(42,30){\color{white}\begin{rotate}{90}shock region\end{rotate}}

\put(10,30){\color{white}\begin{rotate}{90}source region\end{rotate}}
\put(33,30){\color{white}\begin{rotate}{90}source region\end{rotate}}

\put(10.6,28){\color{white}$\rarrowfill{2.2cm}\larrowfill{1.5cm}$}
\put(33,28){\color{white}$\rarrowfill{1.5cm}\larrowfill{0.8cm}$}

\put(79,30){\color{white}\begin{rotate}{90}shock region\end{rotate}}

\put(61,30){\color{white}\begin{rotate}{90}source region\end{rotate}}

\put(61.3,28){\color{white}$\rarrowfill{2.99cm}\larrowfill{2.9cm}$}

\end{overpic}
 \caption{\rev{Wavelet analysis of the extremely large simulation (see Figure \ref{fig:verylarge}) at different times. The color indicates power in arbitrary but consistent units. The arrows indicate the direction in which wavetrains are moving. } \label{fig:wavelet} }
\end{figure*}
In this appendix we present one simulation with an extremely large radial domain, spanning over 55\,000~particle radii.
With the exception of the box size, the simulation parameters are identical to the fiducial simulation (see also Tab.~\ref{table:simulationlist}).
A stroboscopic space-time diagram of the first 6500~orbits is presented in Fig.~\ref{fig:verylarge}.
Once the overstability starts growing three source regions can be identified initially at $t\sim 1000$~orbits.
These are located at $x=-20\,000\,$m, $x=5\,000\,$m and $x=18\,000\,$m.
A few thousand orbits later, $t\sim 4000$~orbits, only two survive.
At the end of the simulation at $t>5000$~orbits, only one source can be seen near $x\sim-26\,000\,$m.
This behaviour is qualitatively identical to the fiducial simulation presented in Sect.~\ref{sec:fiducial}.

\rev{In Figure \ref{fig:wavelet}, we present a wavelet analysis of this simulation using the \texttt{PyWavelets} package\footnote{\url{http://www.pybytes.com/pywavelets/}}.
The plots show the power at a given position and scale at different times.
We present analyses at 3000 and 6000 orbits, which best display the signature of the source and shock structures discussed in Section 4.1.
In both plots, coherent wavetrains are evident in the lines of power concentrated between 350 and 600~m.
In addition, at $t=3000$, sources can be identified at $x\sim 6$~km and $x\sim -25$~km, areas of very low power on all scales.
Shock regions occur between the sources and are indicated as such.
After 6000~orbits, the system has settled down to a state characterised by one source near the left boundary and one shock near the middle of the box (also compare with Figure \ref{fig:verylarge}).
Inbetween, there are wavetrains at a scale of approximately 325~m. 
Note that this scale is smaller than some of the scales observed at earlier times.
This suggests that the system favours a certain scale and will not continue to grow to larger scales indefinitely.
This should not be confused with sources and sinks which are separated by scales much larger than 1~km. 
}

\end{document}